# Co-orchestration of Multiple Instruments to Uncover Structure-Property Relationships in Combinatorial Libraries

*Boris N. Slautin[1]\*, Utkarsh Pratiush[2], Ilia N. Ivanov[3], Yongtao Liu[3], Rohit Pant[4], Xiaohang Zhang[4], Ichiro Takeuchi[4], Maxim A. Ziatdinov[5], and Sergei V. Kalinin[2,5]\**

[1] Independent researcher, 10 Brankova str., 11000 Belgrade, Serbia

[2] Department of Materials Science and Engineering, University of Tennessee, Knoxville, TN 37996, USA

[3] Center for Nanophase Materials Sciences, Oak Ridge National Laboratory, Oak Ridge, TN 37831, USA

[4] Department of Materials Science and Engineering, University of Maryland, College Park, MD 20742, USA

[5] Pacific Northwest National Laboratory, Richland, WA 99354, USA

**ABSTRACT.** The rapid growth of automated and autonomous instrumentations brings forth an opportunity for the co-orchestration of multimodal tools, equipped with multiple sequential detection methods, or several characterization tools to explore identical samples. This can be exemplified by the combinatorial libraries that can be explored in multiple locations by multiple tools simultaneously, or downstream characterization in automated synthesis systems. In the co-orchestration approaches, information gained in one modality should accelerate the discovery of other modalities. Correspondingly, the orchestrating agent should select the measurement modality based on the anticipated knowledge gain and measurement cost. Here, we propose and implement a co-orchestration approach for conducting measurements with complex observables such as spectra or images. The method relies on combining dimensionality reduction by variational autoencoders with representation learning for control over the latent space structure, and integrated into iterative workflow via multi-task Gaussian Processes (GP). This approach further allows for the native incorporation of the system's physics via a probabilistic model as a mean function of the GP. We illustrate this method for different modalities of piezoresponse force microscopy and micro-Raman on combinatorial Sm-BiFeO3 library. However, the framework is general and can be extended to multiple measurement modalities and arbitrary dimensionality of measured signals.

The novel automatized approaches in combinatorial synthesis have revolutionized material design by enhancing cost-effectiveness and significantly accelerating the synthesis rate of new materials.[1–5] Among the oldest examples of high throughput synthesis are combinatorial libraries. More recent examples include pipetting robotics, printing, and many other modalities.[4] However, while the synthesis is by now well established for decades, the bottleneck has been in characterization. Over the last several years automated scanning probe microscopy (SPM), X-ray, Raman microscopy, etc. have opened the way for the readout of structural and functional information from combinatorial libraries and high throughput synthesis that can feed into physical models and materials optimization.

In many cases, of interest are multiple aspects of structure and functionalities across the combinatorial libraries. The investigation of such couplings entails a comprehensive study of multiple electric, mechanical, chemical, and structural material properties. This demands the utilization of a diverse array of local investigative methods such as SPM, electron microscopy, Raman microscopy, and more. Each method serves as a critical tool in unveiling distinct aspects of these couplings, allowing for a thorough understanding across different dimensions and properties. An extensive range of requiring characterization methods often makes the exploration process time-consuming, creating a significant gap between the rates of synthesis and exploration.

In recent years the extensive adoption of machine learning (ML) workflows facilitated the reduction of the substantial gap between synthesis and characterization rates, accelerating the exploration and optimization of targeted material functionalities.[3] The deployed ML agents can provide support in various aspects from mundane data treatment, including image segmentation and features detection[6,7] to big data analysis,[8] automating the decision-making process and governing the experiment itself.[9,10] The concept of automated experiment (AE) in microscopy involves the automatic – *beyond human choice* – operation of the microscope, typically establishing the sequence of imaging or spectroscopy measurement locations. This approach enables operators to redirect their focus from the tedious task of equipment tuning and microscope control at each exploration step towards more advanced result analysis and higher-level decision-making, such as establishing experiment objectives. AE exploration frequently relies on Bayesian optimization (BO), which has proven its efficiency, serving as a robust solution for both material design[1,11–13] and automated material characterization.[14–18] In microscopy, BO was successfully employed for governing the exploration by multiple methods including SPM,[17,19–21] Scanning Transmission Electron Microscopy Electron Energy Loss Spectroscopy (STEM-EELS),[22] 4D STEM,[23] neutron and X-ray scattering.[24–26]

Multiple BO workflows were developed to optimize and accelerate the exploration of the materials.[14,18] Important to note, that employing pure BO can be ineffective for real-world applications. Various modifications have been developed to address these challenges by incorporating additional knowledge. For instance, physics-informed and structures BO inject ground physical models or prior knowledge in the form of a mean function in Gaussian processes (GP).[16,27] The widely used *multi-fidelity* BO methods combine information from different sources, provide various levels of precision, or require different computational resources to efficiently integrate and exploit these datasets to balance accuracy with computational expenses to achieve optimal results.[28,29]

The rapid growth of automated labs brings forth the challenge of the co-orchestrated operation of multiple tools. One example of this can be an exploration of a given materials system with the multimodal tool where, due to technical limitations, data in different modalities can be acquired sequentially. Another example is opened by the combinatorial libraries that can be explored in multiple locations using various tools simultaneously. The third will be similar synthesis systems at different geographic locations with dissimilar downstream analytics. This, in turn, presents a new challenge: developing fast and efficient workflows to navigate the characterization and exploration of the intricate relationship between structure and properties in the arrays of material compositions via synchronous exploration of the identical combinatorial libraries or synthesized samples by multiple tools, balancing the information gain in multiple characterization modalities.

Here we introduce the *multimodal co-orchestration*, representing a significant advancement in optimizing material exploration through the simultaneous orchestration of multiple methodologies. While the workflow is developed for the combinatorial libraries, it can also be used for similar synthesis systems.

**RESULTS AND DISCUSSION**

**Co-orchestration workflow: a concept.**

The alterations in compositions in combinatorial libraries prompt variations in material structure and multiple functional properties, necessitating the utilization of numerous methods (*multimodal discovery*) to study their correlations and intrinsic nature. AE's objective is to uncover the correlation between local composition and target functionality or between dissimilar functionalities in a minimal number of steps. Despite the diverse nature of measured signals, it's anticipated that the property-changing profiles along the compositional change axis in a combinatorial library will exhibit similarity. The essence of multimodal orchestration lies in leveraging knowledge about compositional correlation uncovered for one property to expedite the exploration of another property measured by a different method, thereby accelerating the overall characterization process.

The low-dimensional compositional space within a combinatorial library makes the implementation of the Bayesian optimization for automated discovery a very robust solution. However, different measurements provide specific insights into the system, often resulting in high-dimensional datasets, that render direct application of BO ineffective. The solution lies in encoding the measured data using a variational autoencoder (VAE) to transform the high-dimensional raw data into a low-dimensional representation.[30] As a result, multimodal orchestration occurs by optimizing the exploration trajectory within the low-dimensional space, encompassing composition and VAE latent variables from different modalities.

Here we are providing a delineation of the automated experiment with co-orchestration for two modalities, designated as methods *a* and *b*. To simplify, we also confine our analysis to a case of 1D compositional space of the combinatorial library. However, these principles can be extended and adapted to exploration across a larger compositional space dimensionality and the number of modalities as well. The workflow for co-orchestration can be described as a three-stage process involving 1) seeding stage, 2) initial co-orchestration, and 3) steady co- orchestration (Figure 1).

*1) Seeding stage*

At the preparation stage, the locations $x_{a\,1}, \ldots, x_{a\,n}$ and $x_{b\,1}, \ldots, x_{b\,m}$ for the seed measurements should be chosen for both modalities. The number of predefined positions can be different for various modalities ($m \neq n$). It seems logical to opt for seed positions at the opposite edge of the compositional space, effectively constraining the area of exploration. Thereafter seed measurements performed in the chosen locations result in the multidimensional seed datasets $X_a = [y_a(x_{a\,1}), \ldots, y_a(x_{a\,n})]$, $X_b = [y_b(x_{b\,1}) \ldots y_b(x_{b\,m})]$, – either spectra or images.

*2) Initial co-orchestration*

The received high-dimensional seed datasets are mirrors in low-dimensional latent representation by VAE encoding independently ($X_a \rightarrow Z_a$ and $X_b \rightarrow Z_b$). The dimensionality of the VAE latent space is a hyperparameter that the operator needs to tune. Typically, we are expecting to use only one of the latent variables for subsequent GP learning. However, reflecting high-dimensional data into a 1D latent space makes the latent variable intricate, encompassing both the composition-dependent part and all side and parasitic influences. From this perspective, opting for 2D latent representations increases the degrees of freedom for the VAE. This allows for the potential separation of the target compositional impact from any secondary influences across different latent variables. We select one latent variable from each modality to form a learning dataset for multi-task GP (MTGP). The objective of MTGP is to predict both the mean values ($\bar{z}_{a\,j}, \bar{z}_{b\,j}$) and uncertainties ($\mathbb{V}(\bar{z}_{a\,j}), \mathbb{V}(\bar{z}_{b\,j})$) associated with the selected latent variables (modalities) within the compositional latent space. The multimodal acquisition function is built based on the MTGP outcomes. It guides the selection of the next measurement modality $i$ and location $x_i$ by maximizing/minimizing the acquisition function ($i, x_i = argmax[acq(\bar{z}_{a\,j}, \bar{z}_{b\,j}, \mathbb{V}(\bar{z}_{a\,j}), \mathbb{V}(\bar{z}_{b\,j}))]$). Finally, we augment the seed dataset of the selected modality with the newly acquired data and iterate this initial co-orchestration stage from the beginning.

During the initial co-orchestration stage, the amount of gathered knowledge about the system is limited. Therefore, introducing new data at each exploration step can markedly alter the VAE distributions and the profiles of latent variables throughout the latent space. This mirrors the evolution of our knowledge about the system (i.e. understanding of the relationships between composition and functionality) with acquiring new data. The significant alterations in VAE distributions require retraining the VAE and MTGP at each step from scratch.

*3) Steady co-orchestration*

As information about the system accumulates, injecting new data doesn't drastically alter the VAE distribution. The stability of the VAE distribution and consequent steady latent variable profiles across the compositional space allow for the application of incremental training. This means that with each new data addition, the VAE and MTGP models can be updated without starting training from scratch because the acquired data only augments the existing knowledge of the system rather than revolutionizes it. This incremental learning is associated with the final steady co-orchestration experiment stage.

It is important to note that the transition towards a steady co-orchestration happens independently for different modalities at different times. The number of exploration steps needed for this transition indicates the complexities of composition-dependent functionalities measured by specific modality (method).

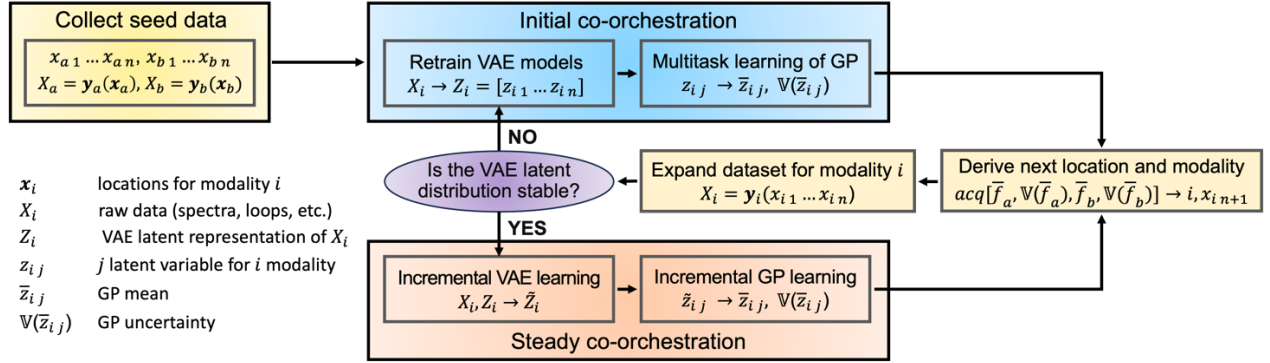

**Figure 1.** Working schematics of multimodal co-orchestration.

### Multi-task Gaussian Processes

The multi-task Gaussian Processes form the foundation for multi-task Bayesian optimization and therefore for multimodal automated experiments as well.[31–33] MTGP extends the concept of GP to handle multiple related tasks, representing distinct objectives or output variables, simultaneously. The model uncovers the dependencies and correlations between tasks, allowing them to leverage information from one task to improve predictions in another. The overall covariance between the outputs in the most general case can be defined as:[33]

$$K[f_a(x), f_b(x')] = K_{ab}^f k(x, x'), \qquad (1)$$

where the $K_{ab}^f$ specifies the inter-task similarities and $k(x, x')$ is the covariance over inputs.

In practice, the Linear Model of Coregionalization (LMC) is often used to capture the correlation between multiple outputs in MTGP. LMC introduces the several ($Q$) shared latent processes with their coregionalization matrixes, denoted as $B^{(q)}$, to establish a linear relationship between multiple tasks. In the LMC-based multi-task GP, the covariance between the $i$-th data point of $z_a$ and the $j$-th data point of the $z_b$ can be formulated as follows:

$$K[z_a(x_i), z_b(x_j)] = \sum_{q=1}^{Q} B_{ab}^{(q)} k_q(x_i, x_j) \qquad (2)$$

In the expression above the $Q$ is the number of the latent processes, $B_{ab}^{(q)}$ represents an element of the coregularization matrix for $q$-th latent process, $k_q(x_i, x_j)$ is the covariance function for $q$-th latent process. The non-diagonal elements of the $B^{(q)}$ are responsible for the multi-task correlations. To ensure the symmetric and positive semi-definiteness of $B^{(q)}$ it is parameterized as:

$$B^{(q)} = W^{(q)}(W^{(q)})^T + diag(v^{(q)}) \qquad (3)$$

where $W^{(q)}$ is low-rank $D \times R$ matrix, $D$ number of tasks (in our case $D = 2$), $R$ is rank, $diag(v^{(q)})$ – encapsulating specific variances for each output. The elements of $W^{(q)}$, and $v^{(q)}$ are estimated directly throughout the learning of GP.

**Co-orchestration experiments**

We employ the Sm-doped BiFeO$_3$ (Sm-BFO) combinatorial library as a model system to demonstrate capabilities of the *multimodal* exploration. This system possesses transition with an increase of Sm content from the ferroelectric state of pure rhombohedral BiFeO$_3$ to a non-ferroelectric state of orthorhombic 20% Sm-doped BiFeO$_3$.[34] The complex nanoscale structural ordering observed for some intermediate states may be a cause of the enhanced electromechanical responses, making compositional dependencies more convoluted.[35,36] Previously, we explored the properties of this combinatorial library by high-resolution STEM.[37,38] Moreover, Sm-BFO was employed to showcase the capabilities of the hypothesis-learning Bayesian optimization workflow.[16] For our experiment simulation, we have utilized two predefined datasets showcasing local electromechanical hysteresis loops and local Raman spectra both measured for identical compositions within the Sm-BFO combinatorial library. Both datasets consist of 94 spectra (hysteresis loops), collected at equidistant intervals along the axis of a compositional gradient. The local composition at the measured locations changes from 20% Sm-BFO at location 0 to pure BFO at location 93.

In the simulations of AE, three distinct modalities were utilized: out-of-field polarization (represented by $amp \cdot \cos(phase)$) of BEPS hysteresis loops),[39] Raman spectra, and the frequency of electromechanical resonance observed during BEPS hysteresis loop measurements under applied voltage. Despite the simultaneous measurement of the polarization and frequency signals in real SPM experiments, their interdependence is indirect.[40] This allows us to consider them as separate modalities for the AE simulation.

**Linear VAE**

VAE representations play a key role in the co-orchestration of AE. The shape and orientation of the latent distribution dictate how the explored data features are aligned to the specific latent variables. In our experiments, at each exploration step, we analyzed the *current latent distributions* of points available at that specific iteration, along with the *whole latent distributions* derived by encoding the entire dataset using the VAE trained solely on the currently available data. It should be noted, that in the real experiment, the whole latent distributions are unavailable during the exploration. Here however it provides the ground truth behavior of the system.

After a few initial exploration steps, we obtained that the substantial expansion of the VAE training datasets by newly acquired data **does not** result in significant alterations in the configuration of both whole and current latent distributions (Figure 2a-d). The variations between distributions at different exploration steps are constrained by their position, degrees of compression/extension in some directions, and random orientation within the latent space. An observed stability in the general shape of the latent distributions indicates the potential ability to move from the initial co-orchestration to a steady co-orchestration stage. However, the random orientation of the latent distributions at each subsequent step results in a redistribution of encoded features among latent variables. This alters latent variables' dependencies on the composition used to train the MTGP.

The stabilization of the orientation of the latent distribution within the latent space is achieved by introducing the Linear VAE (LVAE), a modified version of vanilla VAE. LVAE incorporates *custom loss* in addition to the standard reconstruction and KL divergence losses. This custom loss is computed as the absolute difference between the first latent variable ($z_1$) and its corresponding position along the compositional axis, normalized within the range of $[0,1]$ ($x'$):

$$custom\ loss = |z_1 - x'|, \quad x' \in [0,1] \quad (4)$$

By introducing the custom loss, we encourage the accumulation of linear dependencies in the $z_1$ latent variable, specifically aiming to normalize $z_1$ within the range of $[0,1]$. This approach stabilizes the orientation of latent distribution within the latent space (Figure 2e-h), enabling a transition toward a steady co-orchestration. Important to note that LVAE does not eliminate distribution random reflections and variations in its expansion degree. As a result, profiles of compositional dependencies for different modalities may also reflect and "stretch" relative to the horizontal axis, preserving their overall shape. Some variability in the orientation also conserves and can be regulated through the adjustment of the custom loss normalization range.

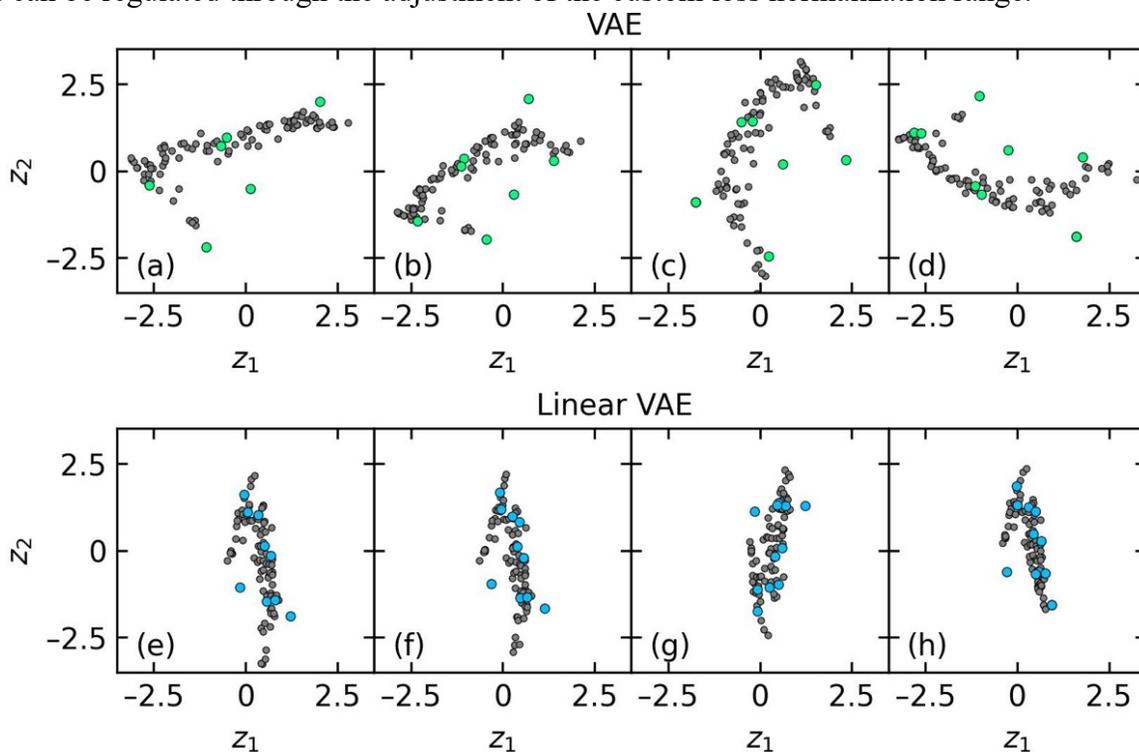

**Figure 2.** Raman spectra latent distributions encoded by (a-d) vanilla VAE and (e-h) LVAE models for the subsequent step in co-orchestration AE. The models were trained from scratch with available data (marked by green and blue points) at each exploration step.

The *ground truth latent distributions* and corresponding compositional dependencies of latent variables were computed for all modalities by training LVAE with the entire datasets (Figure 3).

To evaluate the proficiency of the trained LVAE models in encoding and reconstructing raw data, we decoded spectra from different locations within the compositional profile and compared them with the original raw data. The predicted spectra effectively mirror the dependencies present in the raw data, reducing instrumental noise (Figure S2).

The LVAE fosters the accumulation of features with a proportional relationship between the latent representation and the local composition within the $z_1$ latent variable. Simultaneously, it isolates features with complex compositional dependencies within the $z_2$ latent variable. We utilized latent variables $z_2$ to construct the multimodal space for learning the multi-task GP. Therefore, the primary objective of the co-orchestration AE algorithm is to reconstruct the compositional dependencies (profiles) of the $z_2$ latent variable for each modality, leveraging the correlation between them to expedite the process (Figure 3b,d,f). We observed a predominantly linear downward trend with a few noticeable outlying points in the dependence of $z_2$ on local composition (location) for the Raman measurement (Figure 3b). The $z_2$ values exhibited a growing variability as the composition shifted towards pure BFO. More intriguing dependencies were observed in the latent dependencies of the BEPS modalities. The $z_2$ of BEPS polarization dependence exhibits an extremum at the midpoint of the exploring compositional range (Figure 3d). The downward trend preceding the extremum transitions into a gradual raise in the second part of the profile, accompanied by a noticeable increase in noise levels. This extremum position also aligns with a sharp drop in the BEPS frequency latent variable profile (Figure 3f), where $z_2$ maintains a constant value before this descent. The significant alterations in the middle of the library are associated with the phase transition of the system to the ferroelectric state.

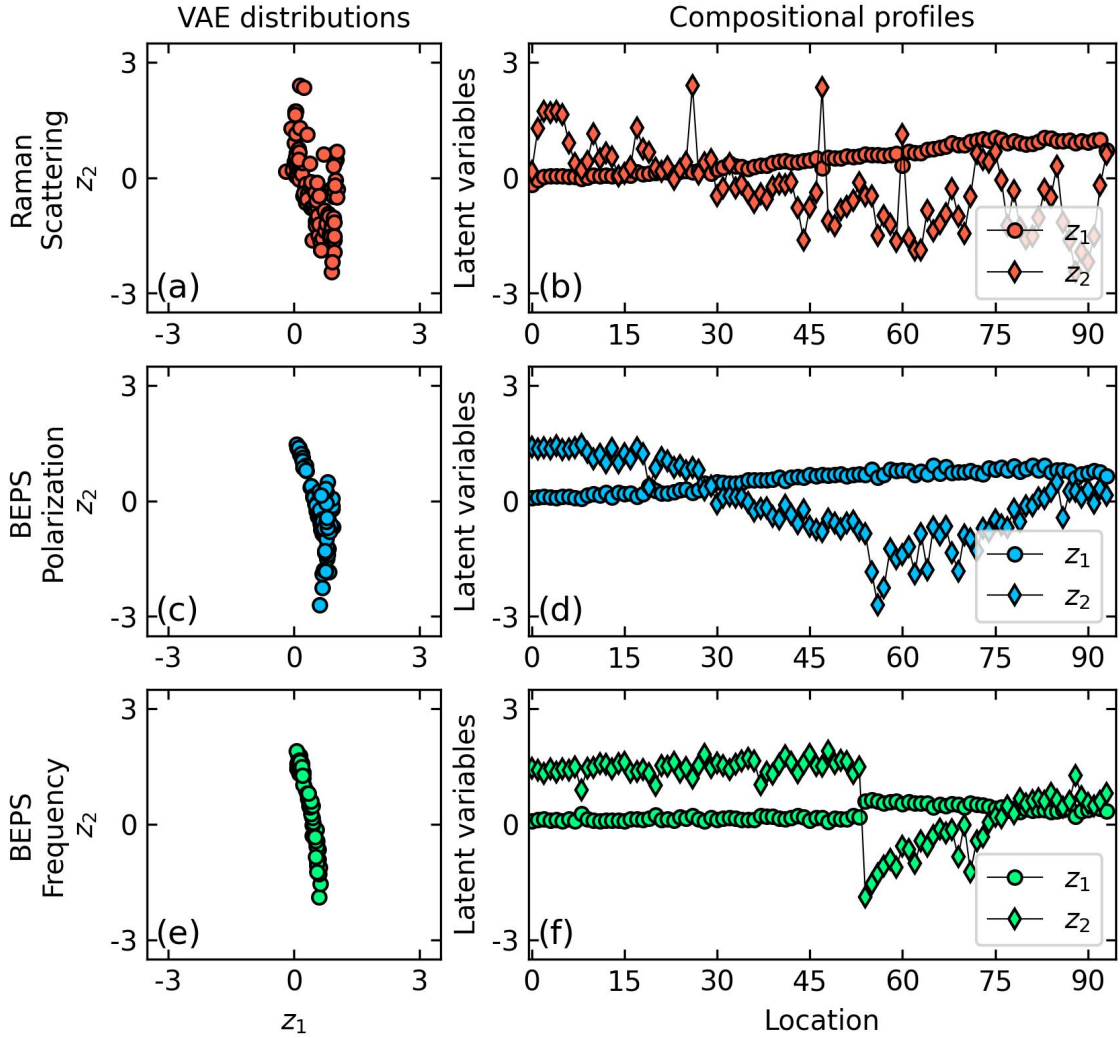

**Figure 3.** LVAE ground truth: (a,c,e) latent distributions and (b,d,f) compositional dependencies of the latent variables for Raman, BEPS polarization, and BEPS frequency modalities.

**Multimodal co-orchestration**

In our simulations, we implement two distinct autonomous experiments. In the *first experiment*, we concurrently explored Raman spectra (Modality 0) and hysteresis loops of BEPS out-of-field polarization (Modality 1) (Figure 4a,c,e,g,i) In the *second experiment*, we co-orchestrate the exploration of out-of-field polarization (Modality 0) along with the BEPS frequency (Modality 1) (Figure 4b,d,f,h,j). The multimodal exploration trajectories were guided by the Maximum Uncertainty acquisition function. In both cases, a total of 30 exploration steps were proceeded. The entire experiments were conducted employing the *initial co-orchestration* approach. For simplicity of analysis, the correlation between the modalities was captured by the MTGP with just two latent processes. Important to note, that the number of latent processes is a hyperparameter in multi-task learning, exerting a substantial influence on the model's performance in terms of generalization and learning speed.

The LVAE, trained with the data available at each exploration step, was employed to encode the entire dataset, establishing the step-specific ground truth (GT) for the $z_2$ compositional dependence (represented by the dashed lines in Figure 4). In both experiments, the ground truth $z_2$ profiles for both modalities maintain their shapes consistently throughout the entire exploration. However, random latent variable profile reflections relative to the horizontal line were observed. Towards the end of the experiment, we observed quite accurate GP predictions for both modalities in the first experiment and specifically for Modality 1 in the second experiment. The reconstruction of the latent variable profile for Modality 0 in the second experiment was hindered by the presence of a sharp drop. For effective reconstruction, it is essential to employ a structural GP and introduce a suitable mean function.

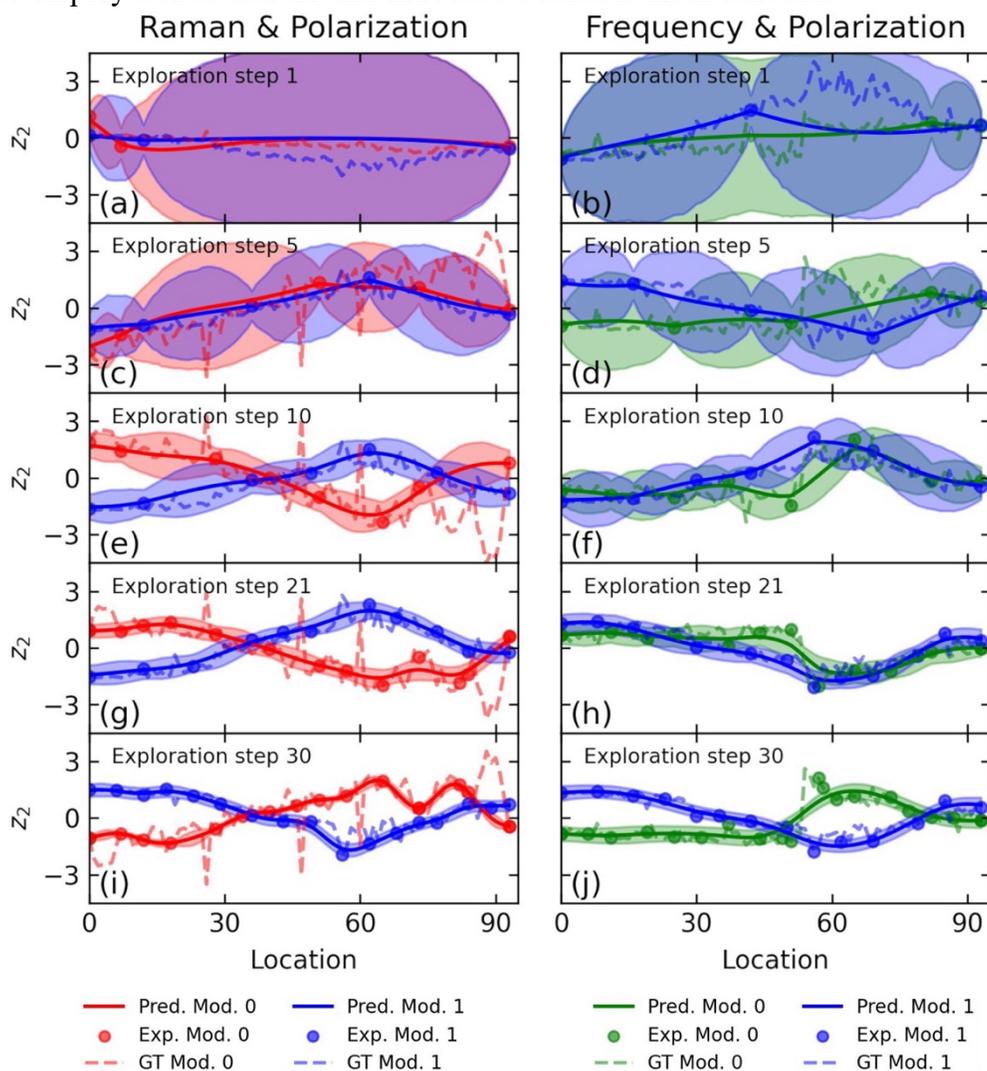

**Figure 4.** MTGP reconstruction at the various exploration steps for (a,c,e,g,i) first and (b,d,f,h,g) second exploration experiments. The GP predictions are represented by solid lines, circles depict the acquired points, and the ground truth (GT) compositional profiles are illustrated by dashed lines.

For a more consistent analysis, we monitor the evolution of the key characteristics of the multi-task BO throughout the entire exploration in both experiments (Figure 5). The experiments exhibit similar behavioral patterns but with some noticeable differences from each other.

The overall downward trends in kernel lengths, $B_{01}$ coefficients of coregularization matrices, prior noise, and GP uncertainties are consistent and preserved in both experiments. The gradual reduction of GP uncertainties, asymptotically approaching zero as the experiments advance, reflects the exploration process (Figure 5g,h). Interestingly, the rate of decrease in GP uncertainties is equal for both modalities. The incremental reduction in $B_{10}$ provides evidence that multi-task coupling plays a significant role at the beginning of the experiment, and its importance diminishes progressively with ongoing exploration (Figure 5c,d). The noticeable jumps in the $B_{10}$ coefficients during the final exploration steps, particularly evident in the second experiment, correspond to the occasional appearance of compositional profile reflections relative to the horizontal line (Figure S3). The prior noise for Modality 0 (BEPS frequency) in the second experiment exhibits higher values and more noticeable variability (Figure 5f). This peculiarity aligns with the sharp drop of $z_2$ value in the middle of the compositional profile. To reconstruct such dependencies, structural GP is commonly employed, whereas the implementation of the vanilla GP (including the MTGP used here) demonstrates limited effectiveness.

We observed the gradual uprising trend in kernel lengths of latent processes at the beginning of AE with subsequent saturation in both experiments (Figure 5a,b). However, during the final steps of exploration in the first experiment, there was a noticeable reduction in kernel lengths (Figure 5a), which was accompanied by a gradual decrease to zero of $B_{10}$ for both latent processes (Figure 5c). We interpret this phenomenon as evidence of the existence of two subsequent exploration phases in the first experiment. During the main part of the exploration, the model endeavors to capture the general trends in compositional profiles of modalities and correlations between them. This leads to the gradual adjustment and stabilization of both the coregularization coefficients and the kernel lengths of the latent processes, consequently facilitating a smooth decrease in the GP uncertainties and prior noises (Figure 5e,g). Once general trends are discovered, the algorithm transitions to the second phase, where the noise characteristics of the profiles are explored, resulting in the reduction of kernel lengths and a decrease to zero in coupling coefficients. The transition to the second phase may be regarded as the trigger to conclude the exploration. We speculate that, in the second simulation, the process did not reach the point of transitioning to noise exploration.

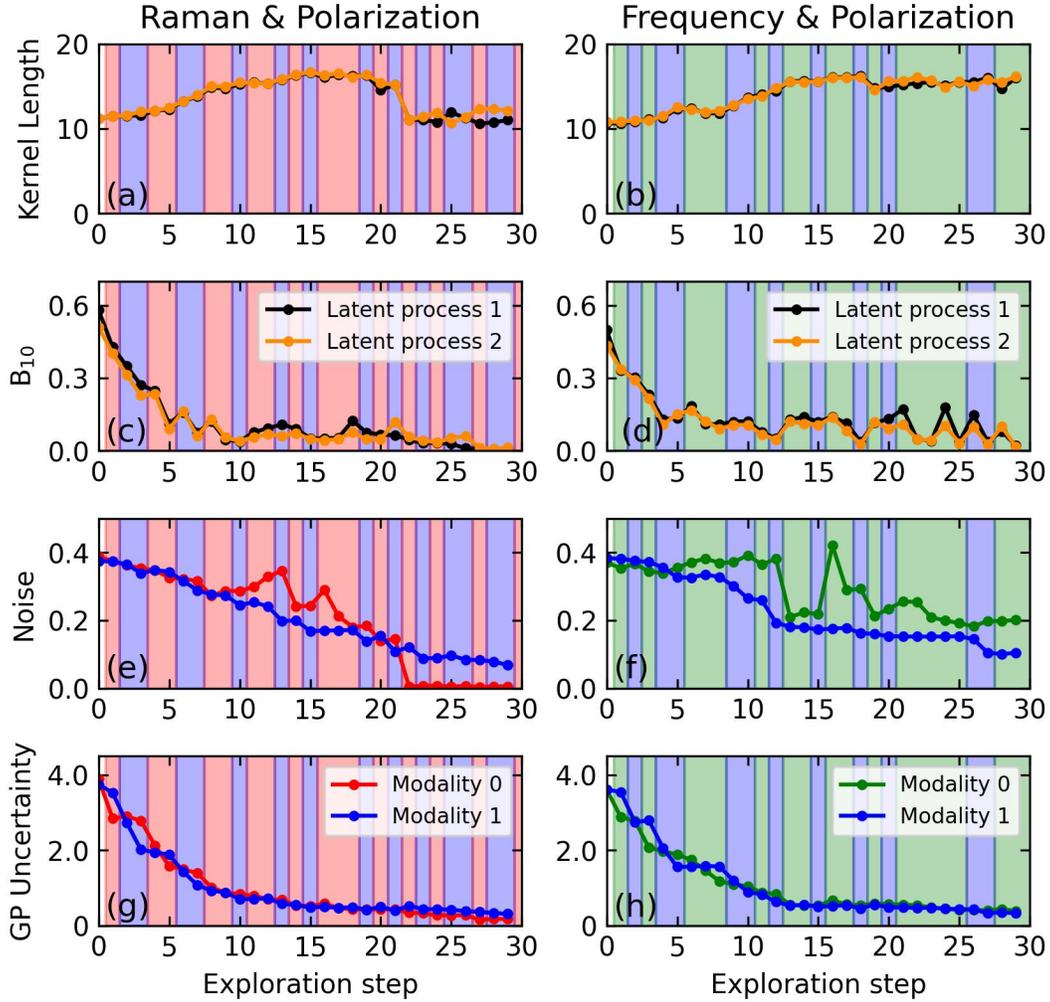

**Figure 5.** MTGP parameters for AE: (a,b) kernel length, (c,d) coregularization coefficient between modalities, (e,f) prior noise, (g,h) GP uncertainty for (a,c,e,g) first and (b,d,f,h) second experiment. The background color strips indicate the selected modality for spectral measurements at each exploration step.

In the final stage of our investigation, we have delved into the possibilities of identifying the point at which the shift from the initial co-orchestration to a steady co-orchestration state becomes feasible. The Kolmogorov-Smirnov (KS) criterion, which measures the maximum difference in cumulative distribution functions, was utilized to evaluate the divergence between the distributions of VAE latent variables and estimate their stability throughout the exploration. At each step, we encode the entire distributions using the VAE, trained on the available points. After that, we calculate the ground truth KS (GT KS) criteria between the encoded distribution and the ground truth latent distribution for the corresponding modality **(Figure 6, diamonds)**. Moreover, KS criteria were also computed at each step to compare a specific latent distribution (represented by only available points) with the distribution from the previous step (Figure 6, circles). We calculated the criterion for each iteration exclusively for the modality to which a data point was added during that specific step. It should be noted, that in the real AE, only the KS criterion can be defined, while GT KS is available only in our simulation.

The KS and GT KS criteria presumably indicate a downward trend, except for the final two steps in Experiment 1, in the latent distributions of the linearized $z_1$ variable (Figure 6). However, the KS criteria for the $z_2$ latent distribution exhibited distinct behavior in the first and second experiments. In Experiment 1, when the latent distributions were stable and exhibited minimal changes with ongoing simulation, we observed the stabilization of the KS and GT KS criteria around 0.2 (Figure 6a). The steady and small values of the KS criterion throughout the exploration can be considered indicative triggers for transitioning towards a stable co-orchestration. Oppositely, in the second experiment, the instability of the latent distribution of Modality 1 (BEPS frequency) results in elevated values observed in both the KS and GT KS criteria (Figure 6b). Developing approaches to enhance the stability of the latent distribution, particularly by mitigating random reflections, is an important future task.

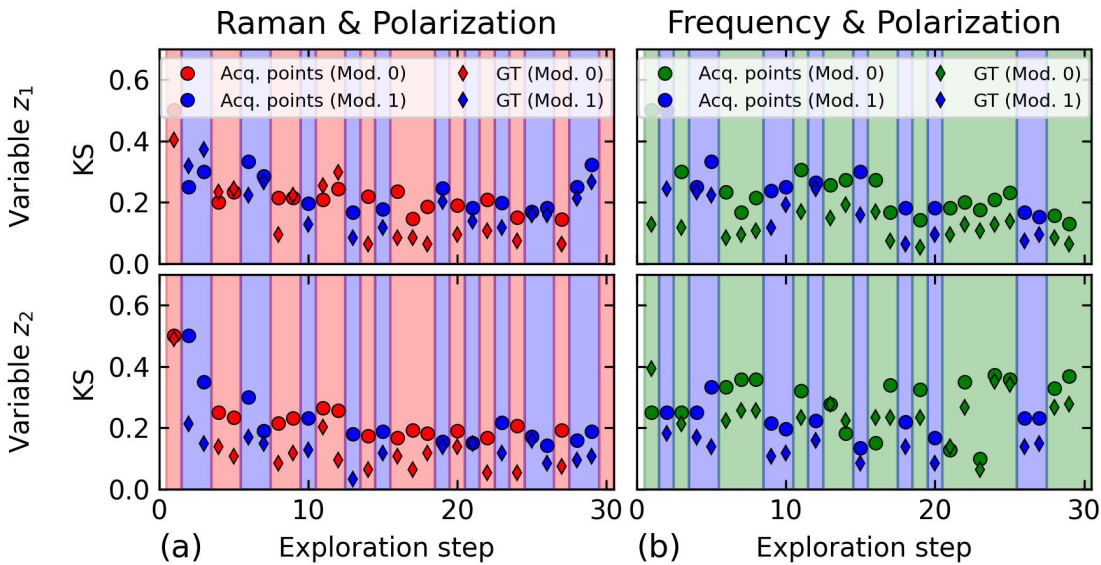

**Figure 6.** Kolmogorov-Smirnov criterion evolution defined for $z_1$ and $z_2$ latent variables for both (a) first and (b) second experiments.

**Conclusions.**

In summary, we have introduced a robust co-orchestration workflow designed to guide the exploration of structure-property relationships in combinatorial libraries through the simultaneous application of multiple methods. The proposed approach is driven by multimodal Bayesian optimization, outlining the optimal exploration trajectory in the compositional space comprising the low-dimensional representations of the raw spectra acquired by different methods. The key advantage of this co-orchestration workflow lies in the real-time utilization of acquired knowledge about the compositional dependency of one property to accelerate the exploration of other properties measured by different methods. This expedites the overall characterization process.

We proposed the utilization of variational autoencoders to encode raw spectra into low-dimensional representations. To improve the orientational stability of the VAE latent distributions throughout the experiment, we introduced the Linear VAE model. In the LVAE, a special *custom loss* was incorporated alongside the standard reconstruction and KL divergence losses during model training to stabilize resulted representations. Ensuring the stability of the latent distribution throughout the experiment enables the acquisition of steady compositional dependencies of the latent variables. The stability of compositional profiles of latent variables, in turn, makes possible transition from the initial co-orchestration stage, where the VAE and MTGP are trained from scratch, toward a steady co-orchestration characterized by incremental learning.

The capabilities of multimodal co-orchestration were validated by the autonomous experiment simulations in the Sm-BFO combinatorial library. The workflow highlights its effectiveness in optimizing the exploration trajectory especially when latent variables exhibit smooth compositional dependencies. Nevertheless, partial effectiveness was also demonstrated in multimodal co-orchestration with a piecewise compositional dependency. The KS criterion, computed for latent distributions at the subsequent exploration steps, is proposed as an indicative signal to determine the possibility of transitioning from the initial to steady co-orchestration stages.

The co-orchestration workflow is expected to significantly enhance the efficiency of combinatorial library exploration, thereby narrowing the gap between synthesis and characterization rates. Further potential improvements could encompass the development of advanced methods for stabilizing VAE representations, the incorporation of structured GP, and the implementation of cost-aware policies for multimodal Bayesian optimization. We do believe that the proposed concept marks the beginning of a new chapter in autonomous experimentation.

**EXPERIMENTAL SECTION**

The exploring combinatorial library is structured in the form of the epitaxial $BiFeO_3$ film synthesized on a (001) $SrTiO_3$ (STO) substrate, where the amount of Sm substitution for Bi is continuously varied from 0 to 20% across the substrate.[37]

The measurement of electromechanical hysteresis loops is conducted using the Band-Excitation Switching Spectroscopy Piezoelectric Force Microscopy approach (BEPS).[41] The loops appear as a response to the modulated triangle waveform applied by the scanning probe microscope's tip. This approach enables the study of the local ferroelectric switching dynamic.[39,42] The measurements were performed within a rectangular grid of locations for better statistics. Following this, the data were averaged along the constant composition axis, resulting in a dataset comprising 94 averaged loops measured along the Sm concentration gradient axis.

The Raman experiments were performed utilizing a 633 nm wavelength laser, which was focused using a 20X lens. This configuration offered a spatial resolution of approximately 5.54 μm. The sampling estimated optical depth is ~134 nm ($n_{BFO} = 2.6$). We normalized the raw Raman spectra by area to exclude the influence of the focus position (Figure S1b). The spectra primarily capture the response from the bulk STO, while the contribution from Sm-BFO epitaxial film constitutes only a minor portion of the overall signal. We extracted a BFO proxy signal by taking the difference between the local Raman spectra and the mean spectrum across the entire dataset (Figure S1c). This enabled us to eliminate the influence of the STO bulk substrate from the collected response. Important to note that the resulting footprint spectra should not be directly interpreted as BFO Raman spectra. However, the observed changes in them, occurring with compositional alterations, are associated with the structural variation within the Sm-BFO compositional library. The spectra were also collected at locations within the square grid and averaged along the constant composition axis. The resulting dataset comprises spectra collected for similar Sm concentrations as those used for the BEPS hysteresis loop measurements.

The GPax Python package was utilized to implement multi-task GP. [43]

## ASSOCIATED CONTENT

**Supporting Information**.

The supporting information file is available free of charge.

The supplementary data includes figures. Figure S1 depicts the preprocessing of Raman spectra, Figure S2 showcases the initial and reconstructed spectra of diverse modalities looped for various Sm-BFO compositions within the combinatorial library, and Figure S3 presents $z_2$ compositional profiles for the subsequent exploration step when LVAE is applied. (PDF)

**Data availability.**

The data that support the findings of this study are available in the supplementary material of this article. Code is available without restrictions at https://github.com/Slautin/2024_Co-orchestration. The GP codes are implemented using GPax package https://github.com/ziatdinovmax/gpax.

## AUTHOR INFORMATION

**Corresponding Authors**
Boris N. Slautin: bslautin@gmail.com
Sergei V. Kalinin: sergei2@utk.edu

**Author Contributions**

The manuscript was written through the contributions of all authors. All authors have given approval to the final version of the manuscript.

## ACKNOWLEDGMENT

The work was supported (S.V.K.) via UTK start-up funding. Confocal Raman characterization was conducted at the Center for Nanophase Materials Sciences (CNMS), which is a US Department of Energy, Office of Science User Facility at Oak Ridge National Laboratory. The work at the University of Maryland was partially supported by ONR MURI N00014172661 and the NIST collaborative agreement 70NANB23H226.

# Supporting Information

Co-orchestration of Multiple Instruments to Uncover Structure-Property Relationships in Combinatorial Libraries

*Boris N. Slautin\*, Utkarsh Pratiush, Ilia N. Ivanov, Yongtao Liu, Rohit Pant, Xiaohang Zhang, Ichiro Takeuchi, Maxim A. Ziatdinov, and Sergei V. Kalinin\**

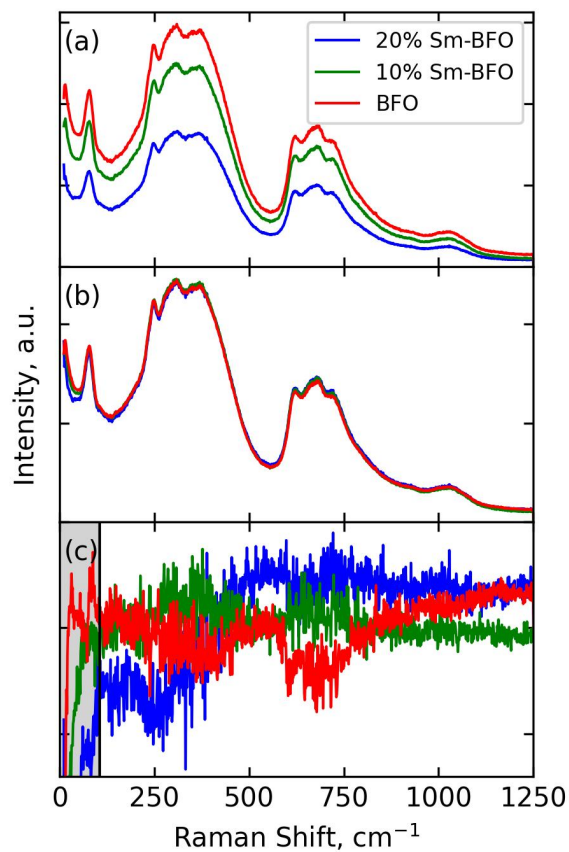

**Figure S1.** The multi-step preprocessing of Raman spectra: (a) raw spectra, (b) spectra after normalization by the area under the curve, (c) footprint spectra, defining as a difference between local spectra and the mean spectrum across the entire dataset. The frequency range highlighted by the grey strip, was removed to diminish the influence of the residual Relay peak wing.

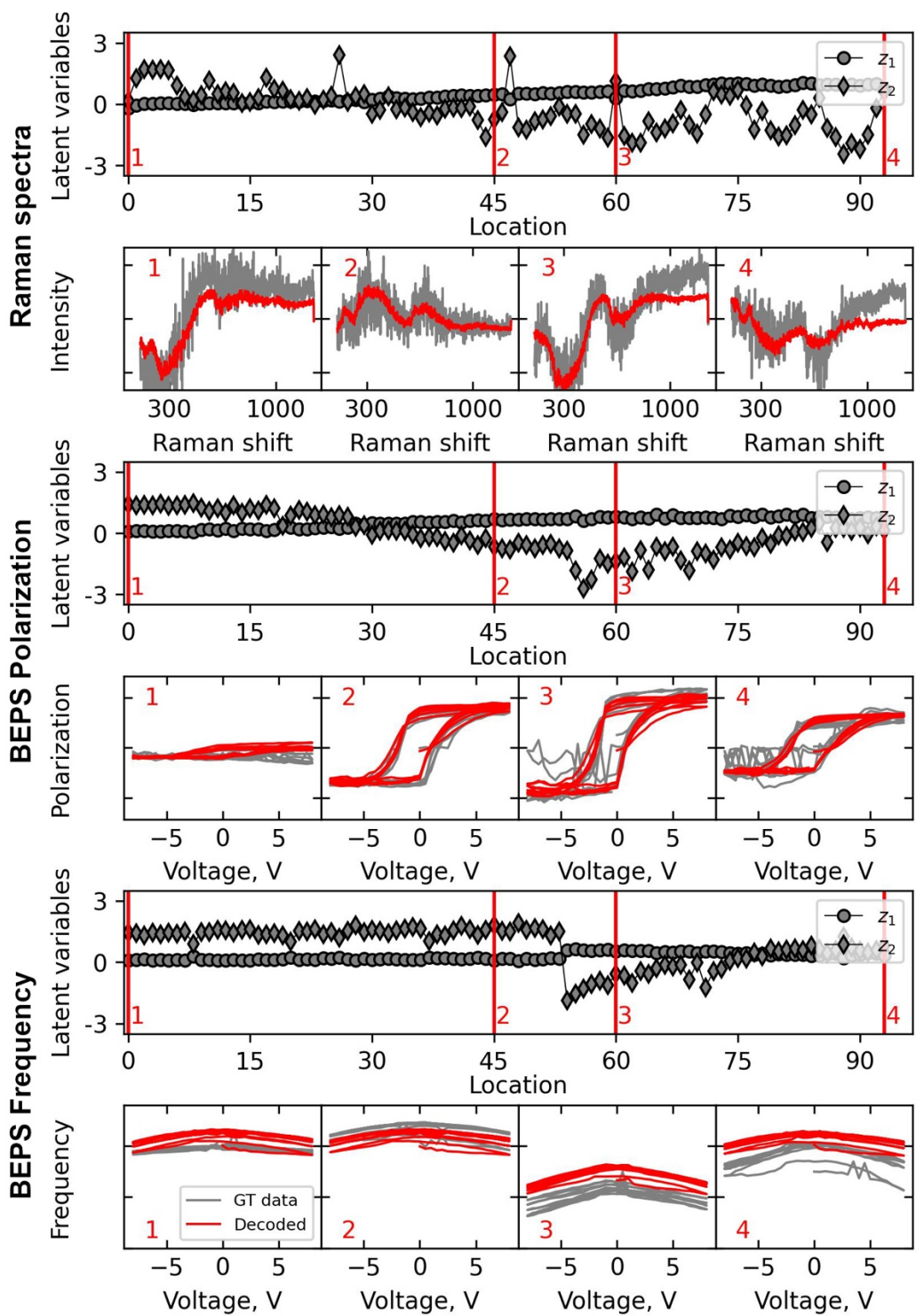

**Figure S2.** Reconstructions of the raw spectra using LVAE for various locations within a combinatorial library. The selected locations are emphasized by red vertical lines. Ground truth (GT) spectra are highlighted in a grey shade, while the LVAE-decoded spectra are depicted in red.

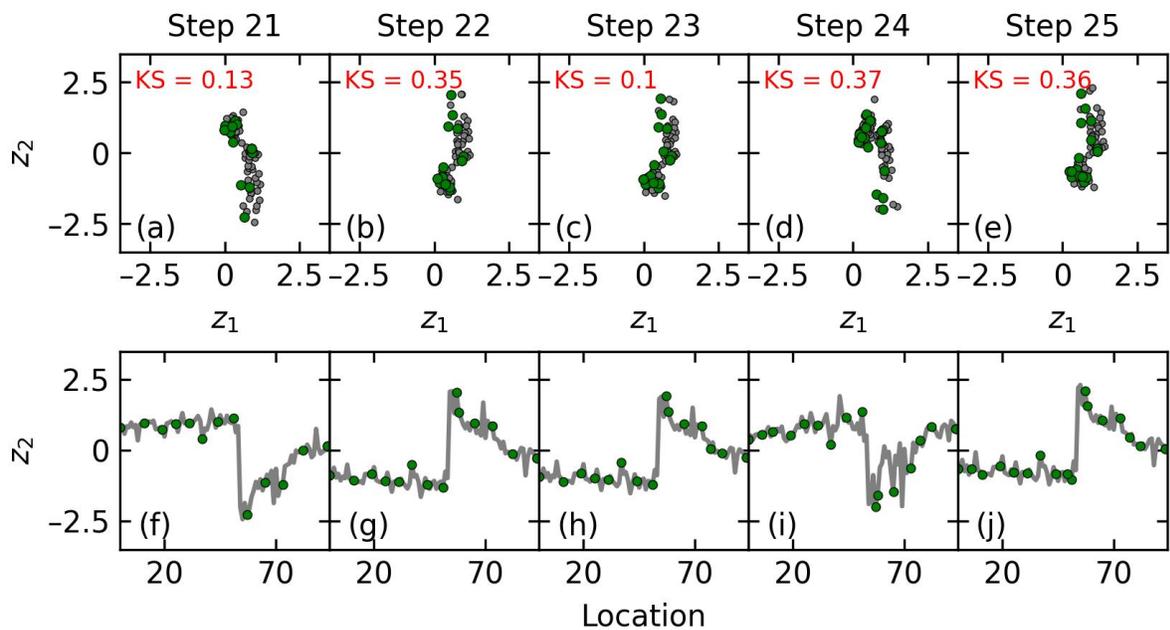

**Figure S3.** (a-e) VAE latent distributions and (f-j) corresponding compositional profiles of $z_2$ latent variables of the Modality 0 (BEPS frequency) for subsequent exploration steps in the second experiment. The KS criteria values in the (a-e) are calculated for pairs – latent distribution in this step and latent distribution in the previous step. The reflections of the distribution in steps 22, 23, and 25 yield high KS values, whereas in step 23, where the distribution orientation is preserved, the KS criterion is minimized.